\documentclass[12p]{iopart}

\usepackage{graphicx}
\usepackage{multirow, array}
\usepackage{color}
\usepackage{xcolor}
\usepackage{url}
\usepackage{float} 

\begin{document}

\title{Benchmarking  {angular-momentum} projected Hartree-Fock as an approximation} 

\author{Stephanie M. Lauber}
\author{Hayden C. Frye}
\author{Calvin W. Johnson}
\ead{cjohnson@sdsu.edu}
\address{San Diego State University,
5500 Campanile Drive, San Diego, CA 92182-1233}


\begin{abstract}
We benchmark angular-momentum projected{-after-variation} Hartree-Fock calculations as an approximation to full 
configuration-interaction results in a shell model basis.  For such a simple approximation we find 
reasonably good agreement between excitation spectra, including for many odd-$A$ and odd-odd nuclides. 
We frequently find shape coexistence, in the form of multiple 
Hartree-Fock minima;  {mixing in shape coexistence, the first step beyond single-reference projected Hartree-Fock}, demonstrably
improves the spectrum in the $sd$- and $pf$-shells.  The complex spectra of germanium isotopes 
present a challenge: for even $A$ the spectra are only moderately good and those of 
odd $A$ bear little resemblance to the configuration-interaction results.  Despite this failure we are able to broadly 
reproduce the odd-even staggering of ground state binding energies, save for germanium isotopes with $N > 40$.
To illustrate potential applications, we compute the spectrum of the 
recently measured dripline nuclide $^{40}$Mg. 
All in all, projected Hartree-Fock  often provides a better description of low-lying nuclear spectra 
than one might expect.  Key to this is the use of gradient descent and unrestricted shapes.
\end{abstract}


\submitto{\jpg}
\maketitle

\section{Introduction}

The excitation spectra of atomic nuclei exhibit a variety of very different behaviors. 
The archetypal patterns of rotational, vibrational, and pairing (or seniority)  are 
 evident in even-even nuclides, and leave their fingerprints on 
 the complex spectra of odd-$A$ and even odd-odd nuclides.  
In response, a variety of methods and approximations have been developed to model atomic 
nuclei~\cite{ring2004nuclear}. 

Of these approaches, mean-field methods such as Hartree-Fock (HF) are the most common starting points, for their 
relative simplicity, appeal to intuition, and flexibility.  Here one has independent particles or 
quasi-particles moving in a field generated by averaging over the rest of the system.  In some 
cases, such as the interacting shell model \cite{bg77,towner1977shell,lawson1980theory,br88,ca05}, the mean-field  fades away to the background: 
the many-body basis states are antisymmetrized products of single-particle states with good 
angular momentum. The interacting shell model, also known as configuration-interaction (CI) is flexible and can easily generate excited states, 
but requires a large  basis to build up correlations. Hence the key to the interacting shell model is to have a simple, efficient representation of the basis allowing for fast 
calculation of matrix elements of the Hamiltonian \cite{Johnson20132761}.  

Alternately, one can use fewer, more complex 
basis states, such as Green's function Monte Carlo \cite{pieper2001quantum} or coupled-cluster calculations \cite{hagen2014coupled}. More closely related to CI are are beyond-mean-field methods such as generator-coordinate methods~\cite{ring2004nuclear,reinhard1987generator,robledo2018mean,klupfel2008systematics} as well as the so-called 
Monte Carlo shell model~\cite{PhysRevLett.77.3315,otsuka2001monte}. These methods also expand the wave function in a basis, but while CI uses 
many simple states to build up correlations,  generator coordinate methods and 
the Monte Carlo shell model uses fewer but more complex states with some  
correlations already built in. In particular, these methods construct mean-field states deformed by external fields, 
either based upon our physics understanding (quadrupole and pairing fields in generator coordinate methods) or sampled stochastically (Monte Carlo shell model). Here conserved symmetries, e.g. rotational invariance and, if using 
quasiparticles, particle number conservation, are broken and must be restored. 
Breaking and subsequently restoring  symmetries builds in important correlations in the wave functions, as well 
as leading naturally to common excitation features such as rotational bands. 

Thus at the heart of beyond-mean-field methods is the projection of broken symmetries ~\cite{ring2004nuclear,jimenez2012projected}, 
 for example angular-momentum projected Hartree-Fock calculations, and generalizations to quasiparticles, number and 
 angular-momentum projected Hartree-Fock-Bogoliubov calculations. 
In this paper we will show that the excitation spectra coming out of 
simple  angular-momentum projected{-after-variation} Hartree-Fock (PHF), 
when compared against full configuration-interaction (FCI)  calculations, are  surprisingly good, not only for strongly deformed 
even-even nuclides, but for a much wider variety of cases, including odd-A and odd-odd nuclides. 
We specifically allow for unrestricted deformation, and find the minima by gradient descent, an important ingredient.
While pairing is known to be an important correlation in nuclei, motivating generalizations to quasiparticles and 
(projected) Hartree-Fock-Bogoliubov (HFB), we investigate neutron separation energies and especially the odd-even staggering (OES) of 
the ground state energy, and find our calculations systematically recover most, though not all, of the OES. 
In many cases we naturally find a second minimum, 
often referred to as 
shape coexistence \cite{RevModPhys.83.1467}; 
 {we go beyond strict single-reference PHF and mix in shape coexistence states,}
which can improve
the excitation spectra. As a particular challenge we consider the  complex spectra of  germanium isotopes. 
We conclude with a calculation of $^{40}$Mg, near the neutron dripline and recently measured 
experimentally~\cite{crawford2019discovery}, and tentatively identify one of the experimentally measured states as a $0^+$ produced by a second minimum.  Thus, although 
going to either FCI or, when impossible, generator coordinate or Monte Carlo shell model methods are often important, 
the very starting point, PHF is already a very good one for describing nuclear excitation spectra.  PHF is also used in quantum chemistry \cite{jimenez2012projected}, another 
motivation for our benchmarking. 

\section{Methods}

In order to compare the projected Hartree-Fock calculations against full configuration 
interaction results, we carry out our calculations in the same shell-model space using the 
same interaction input. 

We have four different spaces with interactions adjusted for 
each one: the $sd$ shell, comprising a frozen $^{16}$O core with valence $1s_{1/2}$-$0d_{3/2}$-$0d_{5/2}$ orbitals, with the universal $sd$ interaction version $B$ (USDB)  \cite{PhysRevC.74.034315}; 
the $pf$ shell, comprising a frozen $^{40}$Ca core with valence $1p_{1/2}$-$1p_{3/2}$-$0f_{5/2}$-$0f_{7/2}$ orbitals, with a modified 
$G$-matrix interaction (GX1A) interaction \cite{PhysRevC.65.061301,PhysRevC.69.034335,honma2005shell}; for our germanium calculations we 
assumed a $^{56}$Ni core and active valence space $1p_{1/2}$-$1p_{3/3}$-$0f_{5/2}$-$0g_{9/2}$, 
and use the JUN45 interaction~\cite{PhysRevC.80.064323}.  Finally, for our calculation of 
$^{40}$Mg we worked in the $sd$-$pf$ space without restrictions, with an $^{16}$O core, and used 
a monopole-modified universal interaction \cite{PhysRevC.86.051301}. 

\subsection{Configuration-interaction}

We benchmark against full configuration-interaction (FCI), sometimes called the interacting shell model \cite{bg77,towner1977shell,lawson1980theory,br88,ca05}, where one expands the wave function in a many-body basis $\{ | \alpha \rangle \}$:
\begin{equation}
| \Psi \rangle  = \sum_\alpha c_\alpha | \alpha \rangle.
\end{equation}
For our basis we use antisymmetrized products of single-particle states, or Slater determinants: 
if $\hat{a}^\dagger_i$ is the creation operator for the $i$th single-particle state, then the occupation representation of 
an $A$-body Slater determinant is
\begin{equation}
\hat{a}^\dagger_1 \hat{a}^\dagger_2 \hat{a}^\dagger_3 \ldots \hat{a}^\dagger_A | 0 \rangle,
\label{SD}
\end{equation}
where $| 0 \rangle$ is the fermionic vacuum, or, equivalently, a frozen core. 

Because both total angular momentum $\hat{J}^2$ and the $z$-component $\hat{J}_z$ commute with 
our Hamiltonians, we choose many-body basis states with fixed eigenvalues of the latter, labeled as $M$. This is known as an \textit{M-scheme basis}, and is easily accomplished when using single-particle states $i$ with good angular momentum $j_i$ and $z$-component
$m_i$, so that the total value of $J_z$ of (\ref{SD}) is $m_1 + m_2 + m_3 + \ldots m_A$.   Other than fixing $J_z$, for FCI 
we take all possible Slater determinants.  In this framework it is easy to construct an orthonormal many-body basis, $\langle \alpha | \beta \rangle = \delta_{\alpha \beta}$. 
The code we use \cite{Johnson20132761,johnson2018bigstick} efficiently computes matrix elements of the Hamiltonian in this basis, $\langle \alpha | \hat{H} | \beta \rangle$, 
and then uses the Lanczos algorithm to find low-lying eigenstates \cite{ca05,Lanczos}.  
It is these FCI results that we use as a benchmark, against which we compare PHF.

\subsection{Hartree-Fock and finding minima}

Our Hartree-Fock code (unpublished but described in \cite{SHERPA}) works in a shell-model space, that is, the single particle states are expanded in an occupation basis: we minimize  
$\langle \hat{H} \rangle $ for an arbitrary Slater determinant $| \Psi \rangle$.  In particular, we redefine the single-particle basis 
by an $N_s \times N_s$ unitary transformation, where $N_s$ is the number of single-particle states,
\begin{equation}
\hat{c}^\dagger_a = \sum_i U_{ia} \hat{a}^\dagger_i,
\label{Utransform}
\end{equation}
 (the only restriction we  impose is $U_{ia}$ is real;  {this is equivalent to an $RT$ symmetry, that is, the wave function is invariant under a rotation by $\pi$ about the $y$-axis, followed 
by time reversal $T$}), and then let 
 \begin{equation} 
 | \Psi \rangle = \hat{c}^\dagger_1 \hat{c}^\dagger_2 \ldots \hat{c}^\dagger_A | 0 \rangle.\label{Slater}
 \end{equation}
Hence the Slater determinant is represented as a rectangular matrix ${\Psi}$, which for $N_p$ particles is given by $N_p$ columns, each of length $N_s$,  of the matrix $\mathbf{U}$ in 
Eq.~(\ref{Utransform}). 
We use separate proton and neutron Slater determinants.
One can compute $\langle \hat{H} \rangle$ for any ${\Psi}$ and then vary the elements of $\mathbf{U}$  to minimize  \cite{SHERPA}.  

One can define local minima by using Thouless' theorem \cite{ring2004nuclear} 
to define an energy landscape in Slater determinants. Because we compare two methods 
of finding minima, and  inquire if these states are indeed minima, we 
review here the relevant details.  Given any reference Slater determinant $| \Psi \rangle$, we  separate single-particle 
states into occupied states labeled by $i,j$ and unoccupied states labeled by $m,n$.
One can then define a particle-hole excitation relative to the reference state, 
$\hat{c}^\dagger_m \hat{c}_i | \Psi \rangle \equiv | m i^{-1} \rangle$. 
The condition for a extremum, and thus a necessary condition for a minimum, 
is the Hartree-Fock condition:
\begin{equation}
    \langle \Psi | \hat{H} | m i^{-1} \rangle = 
    \langle \Psi | \left [ \hat{H}, \hat{c}_m^\dagger \hat{c}_i \right ] | \Psi \rangle = 0.
\end{equation}
Often one introduces the  one-body Hartree-Fock effective Hamiltonian, 
$h_{ab} = \langle \Psi | \left [ \hat{H}, \hat{c}_a^\dagger \hat{c}_b \right ] | \Psi \rangle$, where $a,b$ can be any of $m,i$, and then the Hartree-Fock condition 
is simply stated as $h_{mi} = 0$. An equivalent condition is 
\begin{equation}
    \label{HFcondition}
[ \mathbf{h}, \rho] = 0,
\end{equation}
where 
$\rho_{ab} =\langle \Psi | \hat{c}_a^\dagger \hat{c}_b | \Psi \rangle $ is the 
density matrix, with $\mathbf{\rho} = \mathbf{\Psi \Psi}^\dagger$; $\mathbf{\Psi}$ is the 
matrix representation of the Slater determinant.

 {If $\mathbf{h}$ is diagonal, then the off-diagonal terms $h_{mi}$ must vanish.
Thus, a typical strategy is to diagonalize $h_{ab}$, and use the lowest 
eigenvectors of $\mathbf{h}$ to construct the matrix $\mathbf{\Psi}$.} However 
$\mathbf{h}$ depends upon $| \Psi \rangle$ and hence diagonalization must 
be carried out self-consistently.
 {We iterated diagonalization until we reached convergence in energy; this, however, is not necessary converged in the Slater determinant 
and is not} guaranteed to find an   {energy} minimum: see, for example, in Ref.~\cite{ring2004nuclear} the discussion accompanying Fig.~5.3 how diagonalization can oscillate between two solutions, a situation we often encountered with 
odd numbers of particles.  {Nonetheless, iterated diagonalization is widely used, even in publicly-available and widely used codes \cite{schunck2017solution}.}

To check whether or not one is at an extremum, we check the 
Hartree-Fock condition (\ref{HFcondition}) is satisfied. To check whether one is at 
a local minima, we can find the eigenvalues of the stability matrix \cite{ring2004nuclear},
\begin{equation}
S = \left ( \begin{array}{cc} \mathbf{A} & \mathbf{B} \\ \mathbf{B}^* & \mathbf{A}^* 
\end{array} \right )
\end{equation}
where $A_{mi,nj} = \langle m i^{-1} | \hat{A} | n j^{-1} \rangle$ and 
$B_{mn,ij} = \langle mn i^{-1} j^{-1} | \hat{H} | \Psi \rangle$. The stability 
matrix is related to the famous matrix formulation of the random phase approximation 
(RPA), with a positive-definite stability matrix guaranteeing real RPA frequencies.
As a simpler test, $\mathbf{A}$ is the matrix form of the Tamm-Dancoff 
approximation (TDA), whose eigenstates are 
one-particle, one-hole approximations of excited states \cite{ring2004nuclear,towner1977shell}; the eigenvalues of the TDA must 
also be positive definite for a minimum.

We refer to true local minima, that is, self-consistent solutions that satisfy 
all of our criteria (the Hartree-Fock condition (\ref{HFcondition}) and non-negative 
eigenvalues for the stability matrix--zero eigenvalues arise from degeneracy under rotation) as 
\textit{stable minima}.  In particular  we call the stable global minimum  the \textit{HF minimum}, or the 
primary minimum, and other stable 
local minima as secondary minima.   
While it is not unreasonable to think of  secondary minima as 
examples of shape coexistence \cite{RevModPhys.83.1467}, 
it is possible to find states with the same ``shape'' (i.e. 
quadrupole deformation parameters $\beta$ and $\gamma$, which we find by diagonalizing the mass quadrupole 
tensor as in \cite{PhysRevC.61.034303}), but which are demonstrably different states, 
for example that they have significantly different expectation values of spin or pairing.  Conversely, 
with a pure pairing interaction we  find multiple stable minima with the exact same HF energy but with dramatically different 
shape parameters and even different expectation values of angular momentum $J^2$.  

 {Solutions from diagonalization that are converged in energy but which}
nonetheless fail one or more of our tests 
for a true minimum 
(satisfying the Hartree-Fock condition (\ref{HFcondition}), or either of the TDA or stability 
matrix having negative eigenvalues) we call \textit{unstable solutions}. 
In section \ref{diagVSgrad} we discuss the stability of solutions found self-consistently through diagonalization:  diagonalization  {often but not always}
found stable minima for even-even nuclides, but failed to do so for odd-A and odd-odd nuclides.  Modifications such as 
taking linear combinations of old and new $\mathbf{h}$ did not appreciably change this scenario. 

 {Often one calls upon various measures for odd numbers of particles, such as blocking~\cite{schunck2017solution,PhysRevC.81.024316} or the equal filling approximation~\cite{PhysRevC.78.014304}.
Instead we turn to an alternative to diagonalization:} gradient descent,  \cite{ring2004nuclear,PhysRevC.84.014312}, 
which  more reliably finds stable minima. Here one explores an energy surface by applying
\begin{equation}
| \Psi^\prime \rangle = \exp \left ( -\lambda  \sum_{mi} Z_{mi} \hat{c}^\dagger_m \hat{c}_i  \right )
| \Psi \rangle
\end{equation}
where we let $Z_{mi} = h_{mi}$. The minima we find are generally 
stable (true) minima, 
and always had the same energy or lower than solutions found by diagonalization. 
While secondary stable local minima are common, we only found a small number of saddlepoints, that is, 
solutions which satisfy the Hartree-Fock condition (\ref{HFcondition}) but for which the stability matrix 
has at least one negative eigenvalue.  For example, in the $sd$ shell with the USDB interaction, $^{26}$Ne has a true prolate minimum at -77.26 MeV with $\beta=0.12$, 
but also has an oblate saddlepoint at -76.26 MeV and $\beta = 0.07$. 

As discussed below, the quality of the spectra, i.e., agreement with FCI, was 
generally as good or better with gradient descent minima.
In particular, for odd numbers of particles (odd-odd and odd-A), 
the energies of Slater determinants found by gradient descent were almost always 
significantly lower than those found by diagonalization. 
Although gradient descent 
takes more time to converge to a solution than diagonalization, it was nonetheless 
sufficiently fast we did not need  sophisticated optimization.  
Because with either diagonalization or gradient descent we can rapidly generate many solutions, we generate a large number, typically 10-20, from random starting points; 
in our experience this seems to be sufficient to find all self-consistent solutions.  
 {Increasing the number of random starting points to 1000 did not yield additional solutions.}
Convergence of gradient descent is sped up by doing 10 or 20 pre-diagonalizations.  Because of its 
consistently superior performance, we strongly recommend gradient descent as the 
preferred method of finding minima.

\subsection{Projected Hartree-Fock}

Aside from having good particle number, 
we do not impose any constraints, such as axial symmetry, on our Slater determinants (save, for 
historic reasons, that the matrix representation of the Slater determinant $\mathbf{\Psi}$ be real). 
Allowing arbitrary deformation is important: in general, if one has two minima, 
one with axial symmetry and one without, the triaxial deformed state is usually 
lower in energy.  As a  further example of the importance of  triaxiality, previous work on proton-neutron 
random phase approximation (RPA) of Gamow-Teller transitions ~\cite{PhysRevC.69.024311}, 
found that triaxial HF reference states  better reproduced the FCI 
transition strengths than the reference states with axial symmetry; furthermore, RPA using 
triaxially deformed  Hartree-Fock reference states did a better job than quasi-particle RPA using spherical Hartree-Fock-Bogoliubov reference states, at least in the $pf$ shell. 
Hence allowing arbitrary deformations is a desirable ingredient for our work here.

 An arbitrary state $| \Psi \rangle$, including Slater determinants, can be expanded 
as sum of states with good angular momentum quantum number $J, K$, where $J$ is the total angular momentum and $K$ the $z$-component in the intrinsic frame:
\begin{equation}
| \Psi \rangle = \sum_{J,K} c_{J,K} | JK \rangle.
\end{equation}

Let $\hat{P}^J_{MK} $ be a projection operator that projects out a state of good angular momentum $J$ and $z$-component $K$ in the intrinsic frame, but rotated to $z$ component $M$.
In the standard approach one accomplishes this by an integral  \cite{ring2004nuclear}, but we perform the projection by solving a set of linear algebra equations \cite{PHF1, PHF2}.  Then one 
computes the angular-momentum projected Hamiltonian and overlap kernels, respectively:
\begin{eqnarray}
H^J_{MK} \equiv \langle \Psi | \hat{H} \hat{P}^J_{MK} | \Psi \rangle, \\
N^J_{MK} \equiv \langle \Psi | \hat{P}^J_{MK} | \Psi \rangle, 
\end{eqnarray}
and solves the generalized eigenvalue problem,
\begin{equation}
\sum_K H^J_{MK} g^{(r)}_K = E \sum_K N^J_{MK} g^{(r)}_K.
\end{equation}
where $r$ labels the eigenpairs for a given angular momentum $J$.  {Note that we are doing projection after variation, rather than 
the computationally more intensive variation after projection}. 
These matrices are all of small dimension, and in our model spaces, 
 the HF calculations takes a few seconds and the PHF under a minute on a modest laptop.


\begin{figure}[h!]
    \centering
    \includegraphics[width=\textwidth]{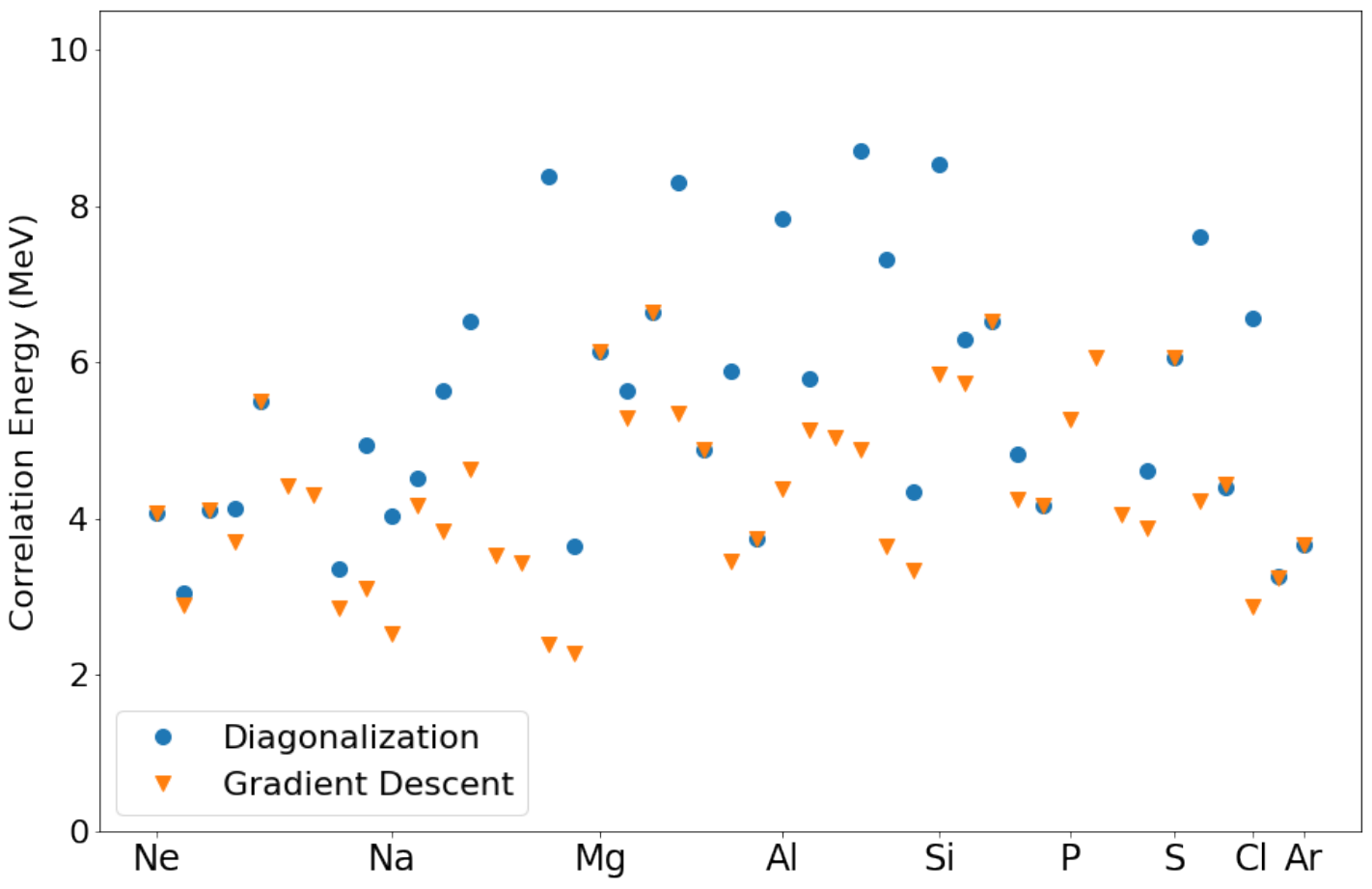}
    \caption{Comparison of finding Hartree-Fock solutions by diagonalization and gradient descent. 
Here we plot the Hartree-Fock correlation energy (the difference between the HF energy and the FCI  ground state energy) for $sd$-shell nuclides, 
using the USDB interaction \cite{PhysRevC.74.034315}.
In all cases where the diagonalization correlation energies are higher, the HF self-consistent solutions are not actual minima.
}
    \label{fig:HF_comparison}
\end{figure}

\section{Results}

\label{results}

Here we present our results.  Note that 
ours are not the first tests of PHF against full CI shell model calculations~\cite{PhysRev.156.1087,whitehead1971exact,whitehead1972shell} 
(there have also been similar benchmarks of PHFB \cite{hammaren1998unrestricted}).  This study is, however, 
the widest systematic benchmarking of PHF against FCI, emphasizing a variety of different nuclides 
including odd-A and odd-odd nuclides. Furthermore, while we obviously cannot reproduce the FCI binding energies with PHF,  in section \ref{separation} we look at 
systematics of the odd-even staggering of the binding energy,  which is expected to be sensitive to pairing.

\subsection{Diagonalization versus gradient descent}

\label{diagVSgrad}

An important question is the relative efficacy of diagonalization and gradient descent finding Hartree-Fock minima.
Fig.~\ref{fig:HF_comparison} compares the HF correlation energies (that is the energy difference between the unprojected Hartree-Fock solution and the FCI ground state energy) for self-consistent solutions found by diagonalization 
and gradient descent,
sample all $sd$-shell nuclides with  $N\geq Z$ and $2 \leq Z \leq 10$ ($^{20}$Ne - $^{36}$Ar). 
Not shown are 8 outliers found by diagonalization with HF correlation energies greater than 10 MeV: $^{25,26}$Ne, $^{26,27}$Na, $^{28}$Al, and $^{30,31,32}$P.

In all even-even $sd$-shell nuclides save $^{26}$Ne, diagonalization and gradient descent found 
equivalent HF minima, that is, the same except for randomly different orientations. (Orientations are found by taking the expectation values of components of angular momentum,
$\langle \hat{J}_x^2 \rangle$, etc..)  
Diagonalization found only unstable solutions 
for $^{26}$Ne, about 6 MeV higher in energy than the true stable minima, a case we have found no explanation for.  
In all  odd-A and odd-odd nuclides, however, gradient descent found a lower, stable minimum, and diagonalization only 
found unstable solutions; 
and all of the unstable solutions found by diagonalization,  save for $^{26}$Ne, are odd-A or odd-odd nuclides. 
Conversely, all solutions found by gradient descent satisfy the HF condition Eq.~(\ref{HFcondition}), and aside from a very 
small number of saddle points all were stable. 
For all $sd$-shell nuclides, the average difference in energy between the (true, stable) HF minimum found by gradient descent, and solutions, stable or unstable found by diagonalization is $\sim 4$ MeV. 

\begin{figure}
    \centering
    \includegraphics[width=\textwidth]{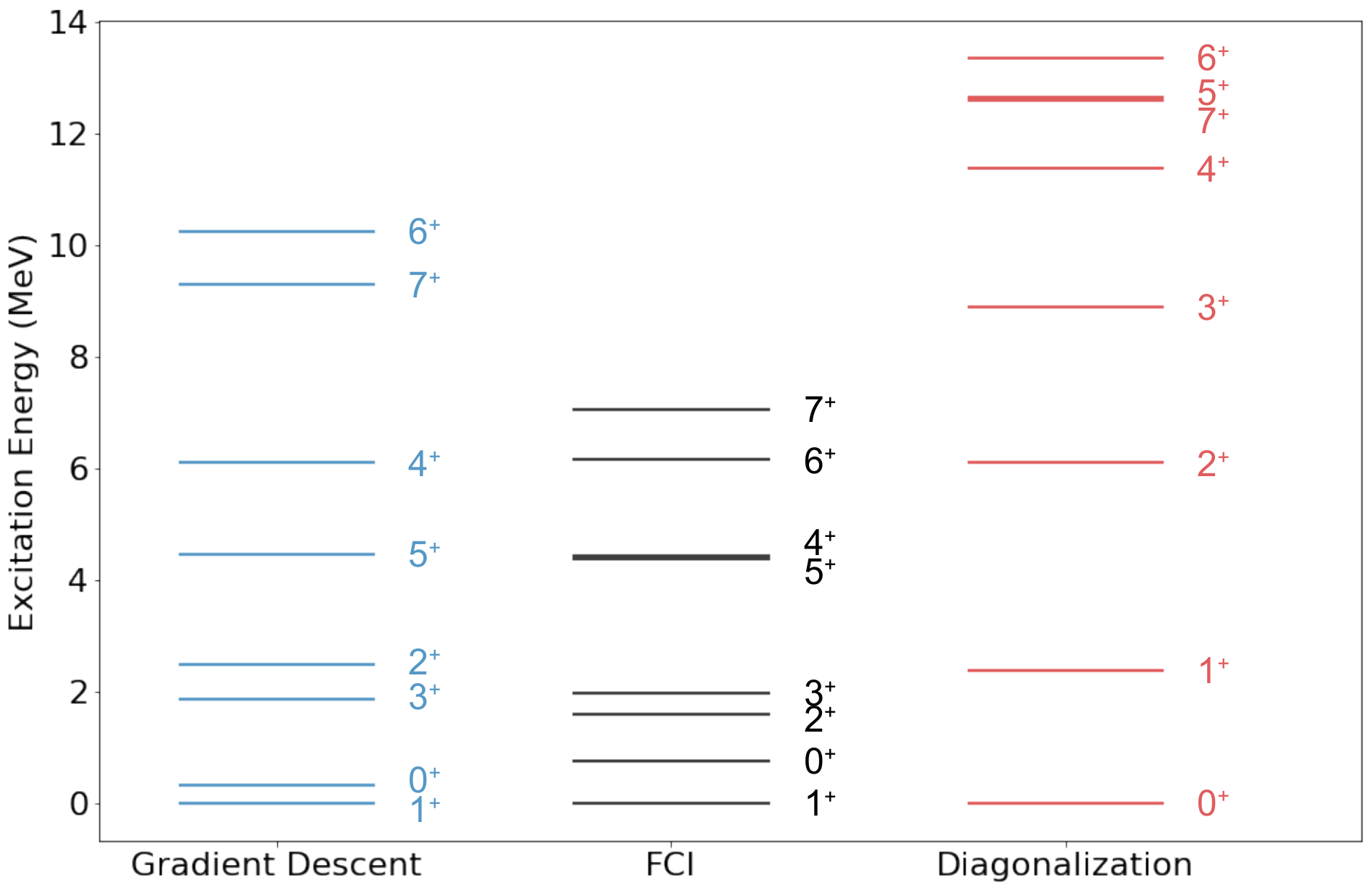}
    \caption{Excitation spectra of $^{30}$P. In the middle is the benchmark 
excitation spectrum from full configuration interaction (FCI), while to the 
left and and right are PHF excitation spectra based upon Hartree-Fock solutions found by gradient descent and diagonalization, respectively.} 
    \label{fig:P30_diagVgd}
\end{figure}

We found a similar result for the $pf$ shell: in only a handful of even-even 
cases (5 out of 45, taking $22 \leq Z \leq N \leq 38$) did diagonalization \textit{not} find the same global minimum 
as gradient descent, and for those five cases the self-consistent diagonalization 
solution was unstable;  while in only five odd-$A$ nuclides did diagonalization 
find the same global minimum as  {gradient descent}. In no odd-odd cases did diagonalization 
find the same stable global minimum as gradient descent.

In addition to lowering the HF energies, gradient descent minima provide better 
PHF excitation spectra than solutions from diagonalization. 
As an example, we provide the excitation spectra for $^{30}$P in 
Fig.~$\ref{fig:P30_diagVgd}$.  
This is a somewhat extreme case:  for diagonalization,   the HF energy is $-122.141$ MeV 
and the PHF g.s. energy is $-126.493$ MeV, while for the gradient descent minimum 
the HF and PHF g.s. energies are  $-148.912$ MeV, and $-151.188$ MeV, respectively.


In the $sd$-shell, gradient descent found stable secondary minima, interpretable as shape coexistence, in about $40\%$ of the cases. (Diagonalization found far more secondary 
self-consistent solutions, but these often did not satisfy the Hartree-Fock condition 
Eq.~(\ref{HFcondition}) and/or were unstable; for the rest of this paper, we will focus 
exclusively on solutions found through gradient descent.) 
In most cases the excitation spectrum was qualitatively improved by the addition of a 
second (local) minimum.  Additionally, mixing in a second minimum lowered the ground state 
energy. The amount of lowering varied strongly: in many cases it was only a few tens of keV, while $^{30}$Si was lowered by 0.63 MeV, $^{46}$Ti by 0.44 Mev, $^{70}$Ge  by 0.73 MeV and $^{71}$Ge by 1.17 MeV.  
For the 21 cases plotted in this paper, the average lowering of the ground state energy of nuclides was 0.22 MeV.  
In the $sd$- and $pf$-shell the average lowering of the ground state energy was 0.16 and 0.18 MeV, respectively, while the germanium ground states were 
lowered by about 0.3 MeV.  The shape parameters $\beta, \gamma$ for the plotted $sd$- and $pf$-shell nuclides 
are found in Table \ref{tab:sdpfshape}, while those for germanium isotopes are in Table \ref{tab:ge}.

\begin{table}[]
    \centering
    \begin{tabular}{|c|r|c|r|r|c|r|}
        \hline Nucleus & $E_\mathrm{HF,1}$  & $\beta_1$ & $\gamma_1$ & $E_\mathrm{HF,2}$  & $\beta_2$ & $\gamma_2$ \\
       & (MeV) &  & (deg) & (MeV) &  & (deg) \\
        \hline $^{20}$Ne & -36.404  & 0.45 & 0  & -31.831  & 0.23 & 60 \\ 
        \hline $^{25}$Mg & -89.113  & 0.23 & 19 & -89.089  & 0.23 & 19 \\
        \hline $^{29}$Na & -104.784 & 0.09 & 0  & -104.692 & 0.09 & 0  \\
        \hline $^{30}$Si & -148.238 & 0.12 & 47 & -148.097 & 0.11 & 60 \\
        \hline $^{30}$Al & -137.851 & 0.10 & 34 & -137.592 & 0.10 & 28 \\ 
        \hline $^{34}$Cl & -199.663 & 0.06 & 60 & -198.554 & 0.06 & 60 \\
        \hline $^{46}$Ti & -67.306  & 0.32 & 0  & -66.474  & 0.24 & 59 \\
        \hline $^{48}$V  & -92.037  & 0.26 & 0  & -91.851  & 0.26 & 0  \\ 
        \hline $^{62}$Ni & -261.307 & 0.12 & 60 & -260.473 & 0.08 & 1  \\
        \hline $^{63}$Zn & -282.779 & 0.11 & 0  & -282.020 & 0.10 & 60 \\
        \hline $^{64}$Cu & -285.099 & 0.10 & 53  & -284.709 & 0.07 & 1  \\
        \hline $^{64}$Ni & -277.062 & 0.07 & 44 & -276.767 & 0.05 & 0  \\
        \hline
    \end{tabular}
    \caption{Hartree-Fock energies, $\beta$ and $\gamma$ deformation parameters for first 2 stable minima for the excitation spectra in Fig. \ref{fig:sdShell} and \ref{fig:pfShell}.  }
    \label{tab:sdpfshape}
\end{table}



In many cases PHF, while a crude approximation, nonetheless reproduces 
excitation spectra, not only 
for rotational spectra as one might naively expect, but in general far better 
than we anticipated.  Naturally adding a second minimum often improves the spectra. 
In this paper we strove to present a general and balanced representation of results, 
 not only the cases where 
PHF and especially the inclusion of a second minimum yields a good result, 
but also less successful cases. 

In the next subsection we  {quantify the goodness of PHF spectra compared to the FCI benchmark spectra}. We then present example excitation spectra from $sd$-shell nuclides, while in section \ref{pfexample} we present examples from the $pf$-shell.  Germanium isotopes 
are known to have triaxial shapes and  complex excitation spectra, with even-$A$ Ge
isotopes typically displaying  
multiple low-lying $0^+$ states from shape coexistence \cite{RevModPhys.83.1467,PhysRevC.25.2812,PhysRevC.76.034317}, and so we investigate 
them as a challenging case in section \ref{Ge}.

Although clearly one would expect the PHF ground state energy to be a poor estimate 
of the absolute binding energy, we nonetheless investigate in section \ref{separation} the relative binding energies 
through neutron separation energies and odd-even staggering of binding energies. 
 {Here one might expect PHF to fail, as the HF solutions lack pairing correlations. We note, however, 
that odd-even staggering has been demonstrated as robust in generic random two-body interactions, even when 
explicit $J=0$ two-body matrix elements, i.e., those associated with pairing, are eliminated~\cite{PhysRevC.61.014311}.
Furthermore, HFB calculations in this region using the same shell-model interactions exhibit only 
weak pairing, especially for well-deformed nuclei \footnote{C.-F. Jiao, private communication}.

}
Once again the results are surprisingly better than we expected.
Only when we get to germanium isotopes with $N > 40$ do we see a significant deviation, 
perhaps signalling the increased role of pairing.

\subsection{ {Goodness of excitation spectra}}
\label{goodness}

\begin{figure}[h!]
    \centering
    \includegraphics[width=0.6\textwidth]{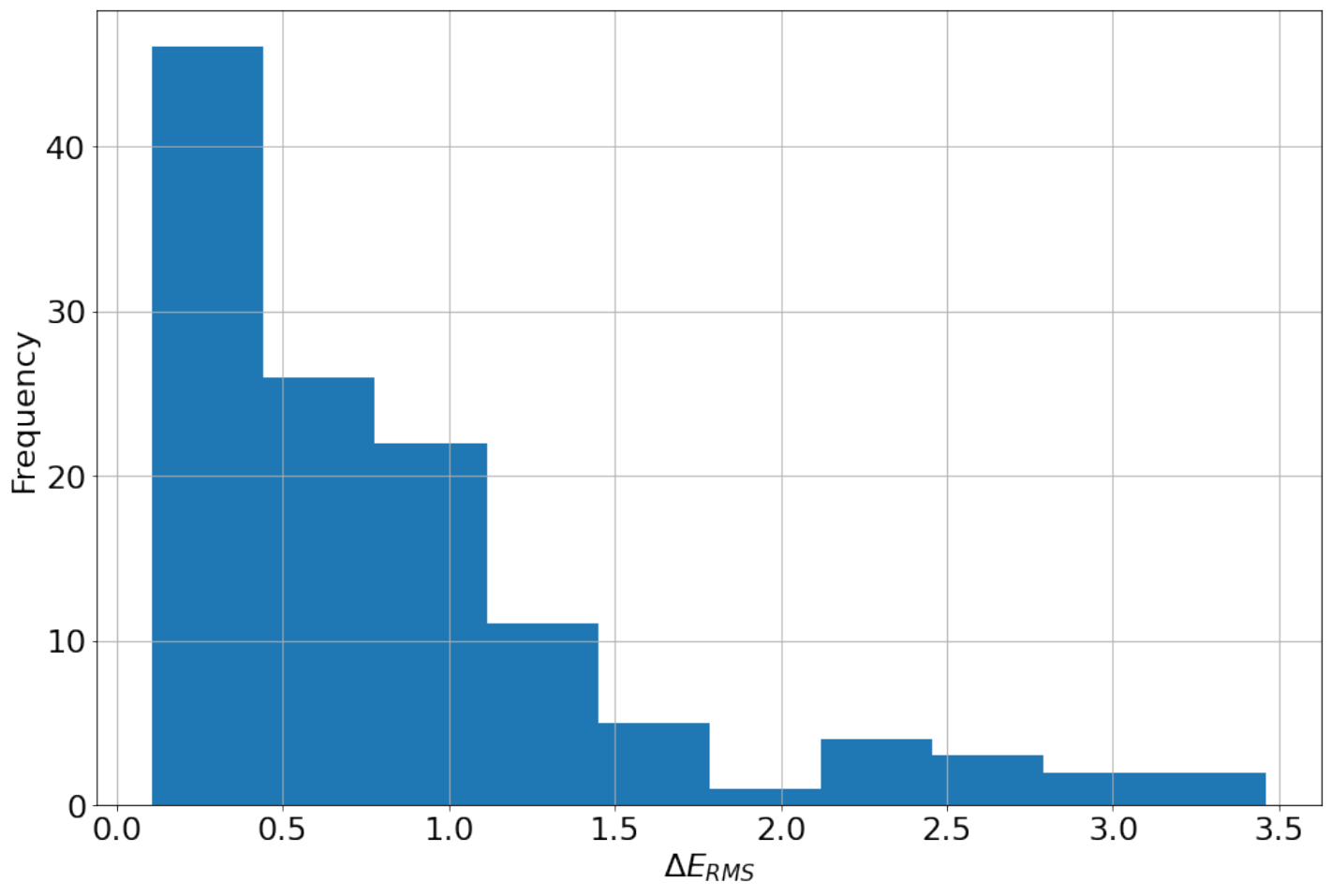}
    \caption{ {The root-mean-square error in the PHF excitation spectra relative to FCI spectra, Eq.~(\ref{dErms}),  calculated for 114 nuclei across the $sd$- and $pf$-shells and 8 even-even germanium isotopes. The average error was $0.85$ MeV for the total set and $0.38$ MeV for germanium. 
    } }
    \label{fig:rmsHistogram}
\end{figure}

 {To approximately quantify the goodness of the PHF excitation spectra, we computed the root-mean-square error 
for $N_\mathrm{lev}$ levels in the excitation spectrum of a given nuclide,
\begin{equation}
\Delta E_\mathrm{RMS}^2 = \frac{1}{N_\mathrm{lev}}\sum_{\alpha=1}^{N_\mathrm{lev}} 
\left(  E_\alpha^\mathrm{FCI} - E_\alpha^\mathrm{PHF} - E_\mathrm{shift} \right )^2.
\label{dErms}
\end{equation}

Here we matched FCI and PHF levels with the same $J$, in order of their appearance. $E_\mathrm{shift}$ is a floating shift that allows for 
different ground state angular momenta or other aspects: for example, in many cases the spacing in PHF between a $0^+$ ground 
state and the first $2^+$ state is significantly smaller than for FCI, while other levels have the approximately correct 
spacing. We found $E_\mathrm{shift}$ by minimizing $\Delta E $.  Because we could not always be certain we correctly  matched 
physically corresponding levels, if the separation between a PHF and matched FCI level was 
greater than $2\Delta E_\mathrm{RMS}$ we excluded 
that pair and recomputed.  While this is far from a rigorous procedure, in nonetheless provides quantitative insight 
into the quality of the PHF spectra. In our figures below we have the PHF spectra shifted by $E_\mathrm{shift}$.
}

 {Fig.~\ref{fig:rmsHistogram} shows
the distribution of $\Delta E_\mathrm{RMS}$ for 114 nuclides across the $sd$- and $pf$-shell and 8 even-$A$ germanium isotopes with more than 2 matched pairs of PHF and FCI energies.
The average error was $0.85$ MeV.  For even-even nuclides the average error was $0.67 $ MeV, for odd-odd nuclides $0.68$, and for odd-$A$ $1.05$ MeV; all had a Poisson-like distribution much as Fig.~\ref{fig:rmsHistogram}. For the germanium isotopes the average error was lower than the total set at $0.38$ MeV; this smaller number is likely due to their compressed energy spectra and the presence of a second HF minima in 7 out of the 8 nuclei investigated. For comparison, the rms error between empirical FCI calculations and 
    experimental spectra is typically on the order of 0.15-0.30 MeV~\cite{br88}.
 }

\begin{figure}[h!]
    \centering
    \includegraphics[width=0.9\textwidth]{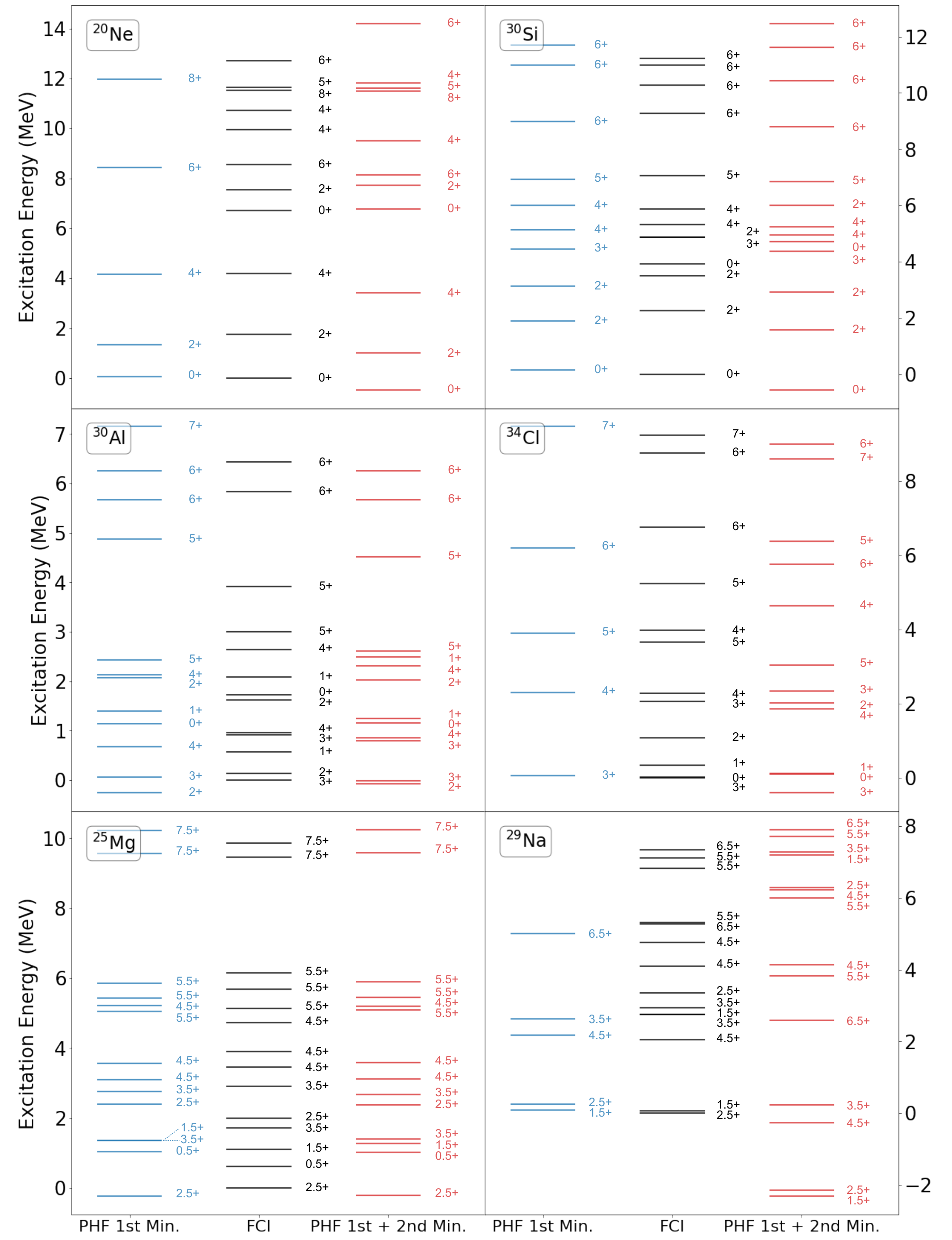}
    \caption{Excitation spectra for selected $sd$-shell nuclides, using the USDB interaction \cite{PhysRevC.74.034315},  computed in FCI (middle spectra), in PHF with  
the primary global HF minimum (left-hand spectra), and PHF mixing the primary and secondary HF minima (right-hand spectra). 
    } 
    \label{fig:sdShell}
\end{figure}

\subsection{Examples from the $sd$-shell}

\label{sdexample}

We chose six illustrative cases in the $sd$-shell, shown in Figure \ref{fig:sdShell}:  two even-even nuclei ($^{20}$Ne and $^{30}$Si), two odd-odd nuclei ($^{30}$Al and $^{34}$Cl) and two odd-A nuclei ($^{25}$Mg and $^{29}$Na). All had stable primary and secondary minima with different Hartree-Fock energies and deformation parameters.  
 {As we allowed the PHF excitation spectra to float in order to miminize $\Delta E_\mathrm{RMS}$, the ground 
states do not completely align. In many cases this allows us to see that the PHF spectra tend to be less compressed 
than the FCI spectra, a phenomenon found not only in the $sd$-shell but also in the $pf$-shell.}

\emph{Even-even.} In both even-even nuclei, the second minimum reproduces a second low lying $0^{+}$ state, 
a second bandhead, at roughly the correct excitation energy. In $^{20}$Ne, the second miminum roughly 
reproduces a second rotational band.
In $^{30}$Si, the  $0^{+}_2$ state is correctly placed between the  $2^{+}_2$ and  $3^{+}_1$ state found in FCI calculations. 

\emph{Odd-odd.}  PHF fails to provide the correct $3^+$ ground state for $^{30}$Al but the addition of the second minimum lowers the $3^{+}_1$ excitation energy by roughly 800 keV, nearly 
degenerate with the $2^+_1$ state. 
Adding the second minimum also puts the $4^+_1$ and $3^+_2$ very close in energy, as in the FCI spectrum. 

A more significant improvement is found  in $^{34}$Cl.   While even PHF with a single minimum produces the correct $3^+$ ground state,  including the second minimum 
produces the  $0^+_1, 1^+_1$ states just above the ground state, as well as 
the $3^+_2$ at near the correct excitation energy. 

\emph{Odd-A.} The addition of the second minimum for $^{25}$Mg lowered the ground state by only 1 keV, the smallest change in the $sd$-shell nuclei in this study.
Here PHF, while roughly reproducing the ordering of low-lying states, clusters them in a way not reflected in the FCI 
spectrum, and addition of the second minimum only mixes minimally. Conversely, while PHF fails (barely) to get 
the correct ground state spin for $^{29}$Na, even the the HF global minimum alone reproduces the tight clustering 
of the $5/2^+_1, 3/2_1^+$ states and of the $7/2_1^+, 9/2_1^+$, while also placing the $11/2_1^+$ and $13/2_1^+$ 
at roughly the right place.


\subsection{ {Examples from} the $pf$-shell}

\label{pfexample}

\begin{figure}[h!]
    \centering
    \includegraphics[width=0.9\textwidth]{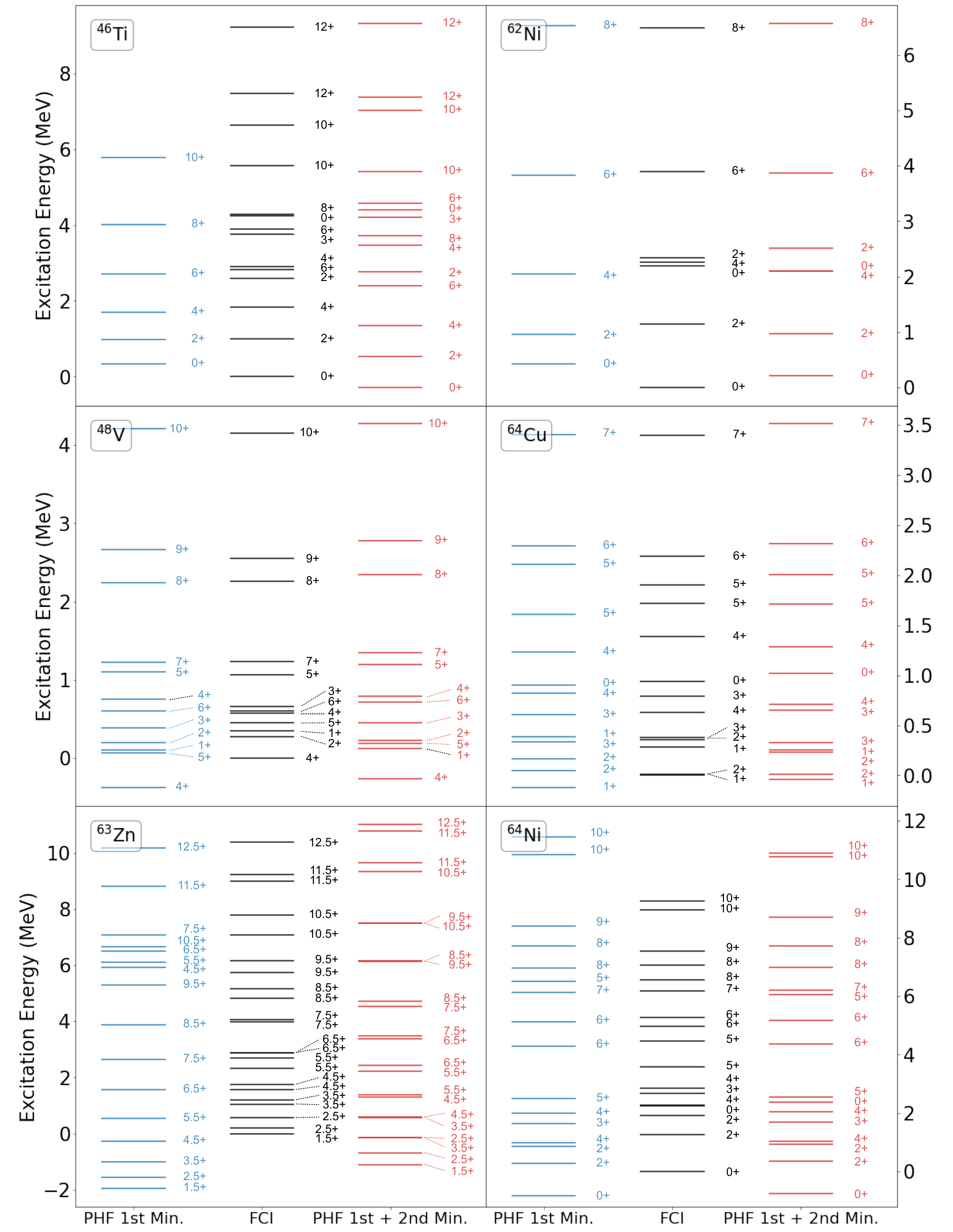}
    \caption{Same as Fig.~\ref{fig:sdShell}, but for selected $pf$-shell nuclides using the GX1A interaction \cite{PhysRevC.65.061301,PhysRevC.69.034335,honma2005shell} .} 
    \label{fig:pfShell}
\end{figure}

The 6 benchmark cases for the $pf$-shell are shown in Figure \ref{fig:pfShell} with three even-even nuclei ($^{46}$Ti and $^{62,64}$Ni), two odd-odd nuclei ($^{48}$V and $^{64}$Cu) and one odd-A nucleus ($^{63}$Zn). Similar to the $sd$-shell results (above) we see improvement in the excitation spectra for the even-even nuclei, modest improvement in the odd-odd nuclei and the case where the addition of a second local minimum produces energy level doublets similar to those seen in FCI. 

\emph{Even-even}. As expected, the even-even nuclei show improvement in their spectra with the addition of the second minimum correctly reproducing a second $0^{+}$ state for both. In $^{46}$Ti, the excitation spectrum from 
just the  prolate global HF minimum provides a mediocre approximation to the exact FCI results, but the inclusion of the second minimum  adds $2^{+}_2$ and $5^{+}_1$, in reversed order but in approximately the right places, with the same for 
the $3^{+}_1, 0^{+}_2, 8^{+}_1$ states. Similar behavior occurs for $^{62}$Ni, which has an approximate vibrational 
spectrum: the spectrum from the  oblate global HF minimum alone is closer to rotational, but mixing in the second local minimum 
dramatically improves this, with the cluster of $0^+_2, 2^+_2, 4^+_1$ in roughly the right place, albeit with the 
$0_2^+$ and $4_1^+$ reversed. 

Conversely, the excitation spectrum of $^{64}$Ni from just the triaxial global HF minimum is already reasonably good,
getting the ordering of the first several states and their spacing approximately correct, save for the $0_2^+$ 
state which only appears  with the addition of a second local minimum.

One trend to note here is the $2_1^+$ excitation energy tends to be systematically low, more so in the $pf$-shell
than in the $sd$-shell. In section \ref{Ge} we will see this trend in germanium isotopes as well. One possible explanation is 
that the $0^+$ ground state is in fact not low enough, due to 
inadequate treatment of pairing. As we will discuss in our conclusions, however, the evidence regarding pairing is mixed.

\emph{Odd-odd}. Even with just the global HF minimum, the low-lying PHF excitation spectrum $^{48}$V matches 
qualitatively the FCI spectrum, getting the correct ground state spin and the low-lying states in approximately the correct order, in particular the near-doublets of $5^+_3, 7^+_1$ and $8_1^+, 9_1^+$, as well as the approximate 
location of the $10^+_1$. Addition of a second local minimum only slightly changes the spectrum.  Conversely, 
$^{64}$Cu is significantly improved by the addition of the second HF minimum, reproducing the ground state near-doublet 
$1_1^+, 2_1^+$ and the $1_2^+, 2_2^+, 3_1^+$ cluster around 300 kev.  The $0_1^+$ state is placed 
correctly near 1 MeV excitation energy.  This is all the most impressive when 
noting the high density of low-lying states.



\emph{Odd-A}. 
The FCI spectrum of $^{63}$Zn is characterized by $5/2^+, 7/2^+, 9/2^+, 11/2^+$ doublets. When projecting from 
only the prolate HF global minimum, we get only one of each doublet, at approximately the correct excitation energy; adding the second, oblate 
local minimum produces the angular momentum partners but not in the correct relative positions.

\subsection{Shape coexistence: Ge isotopes}

\label{Ge}

\begin{table}[h]
\label{cfGe}
    \centering
    \begin{tabular}{|c|c|cc|cc|c|}
\hline
Nuclide & Force & $\beta_1$ & $\gamma_1$ & $\beta_2$ & $\gamma_2$ 
 & $\Delta E_\mathrm{soln}$  \\
 &  &  & (deg) &  & 
(deg) &  (MeV) \\
\hline
$^{64}$Ge & JUN45 & 0.28 & 20  & - & -  & -\\
 & JJ44  & 0.30 & 19 & 0.18 & 60 & 0.36 \\
 & SLy6 & 0.24 & 28 & - & - & -  \\
 \hline
$^{66}$Ge & JUN45 & 0.21 & 60 & 0.24 & 8.1 & 0.58\\
 & JJ44  & 0.25 & 12 & 0.21 & 58 & 0.17 \\
 & SLy6 & 0.23 & 60 & 0.21 & 0.0 & 0.53 \\ 
\hline
$^{68}$Ge & JUN45 & 0.17  & 38  & - & -  &  -\\
 & JJ44  & 0.24 & 40  & 0.13  & 60  & 0.33  \\
 & SLy6 & 0.21 & 39 & -  & - & - \\ 
\hline
$^{70}$Ge & JUN45 & 0.11  & 60  & 0.21 & 36  &  0.08 \\
 & JJ44  & 0.21 & 34  & -  & -  & -  \\
 & SLy6 & 0.23 & 34 & 0.17  & 60 & 1.93 \\ 
\hline
$^{72}$Ge & JUN45 & 0  & -   & 0.16 & 44  &  0.87 \\
 & JJ44  & 0.19 & 31  & 0.02  & soft  & 3.35  \\
 & SLy6 & 0.0 & - & 0.22  & 34 & 0.70 \\ 
\hline
$^{74}$Ge & JUN45 & 0.15  & 41   & 0.06 & 60  &  1.35 \\
 & JJ44  & 0.17 & 18  & 0.15  & 37  & 0.45   \\
 & SLy6 & 0.24 & 28 & 0.07  & 0  & 4.90 \\ 
\hline
$^{76}$Ge & JUN45 & 0.13  & 29   & 0.10 & 29  &  1.23 \\
 & JJ44  & 0.13 & 23  & -  & -  & -   \\
 & SLy6 & 0.16 & 0 & 0.24  & 24  & -2.94  \\ 
\hline
$^{78}$Ge & JUN45 & 0.09  & 12   & 0.10 & 21  &  0.74 \\
 & JJ44  & 0.11 & 25  & -  & -  & -   \\
 & SLy6 & 0.18 & 0 & -  & -  & - \\ 
\hline
    \end{tabular}
    \caption{Comparison of mean-field deformation parameters for even-even germanium isotopes, using 
 two shell-model interactions in the $1p_{1/2, 3/2}$-$0f_{5/2}$-$0g_{9/2}$ 
space with a $^{56}$Ni core, JUN45 \cite{PhysRevC.80.064323} and JJ44 \cite{PhysRevC.70.044314,PhysRevC.76.054312}, as well as Skyrme-BCS calculations 
using the SLy6 parameterization \cite{chabanat1997skyrme} as tabulated in \cite{PhysRevC.76.034317}.  Here $\beta_1, \gamma_1$ and $\beta_2, \gamma_2$ refer 
to the quadrupole shape parameters of the first and, if it exists, second HF minima, 
respectively.  $\Delta E_\mathrm{soln}$ is the energy difference between the 
first and second HF minima if the latter exists.  {`Soft' here means a range of values of $\gamma$ had nearly 
identical energies, within the numerical tolerance of our code.}
}
    \label{tab:ge}
\end{table}

\begin{figure}[h!]
    \centering
    \includegraphics[width=0.9\textwidth]{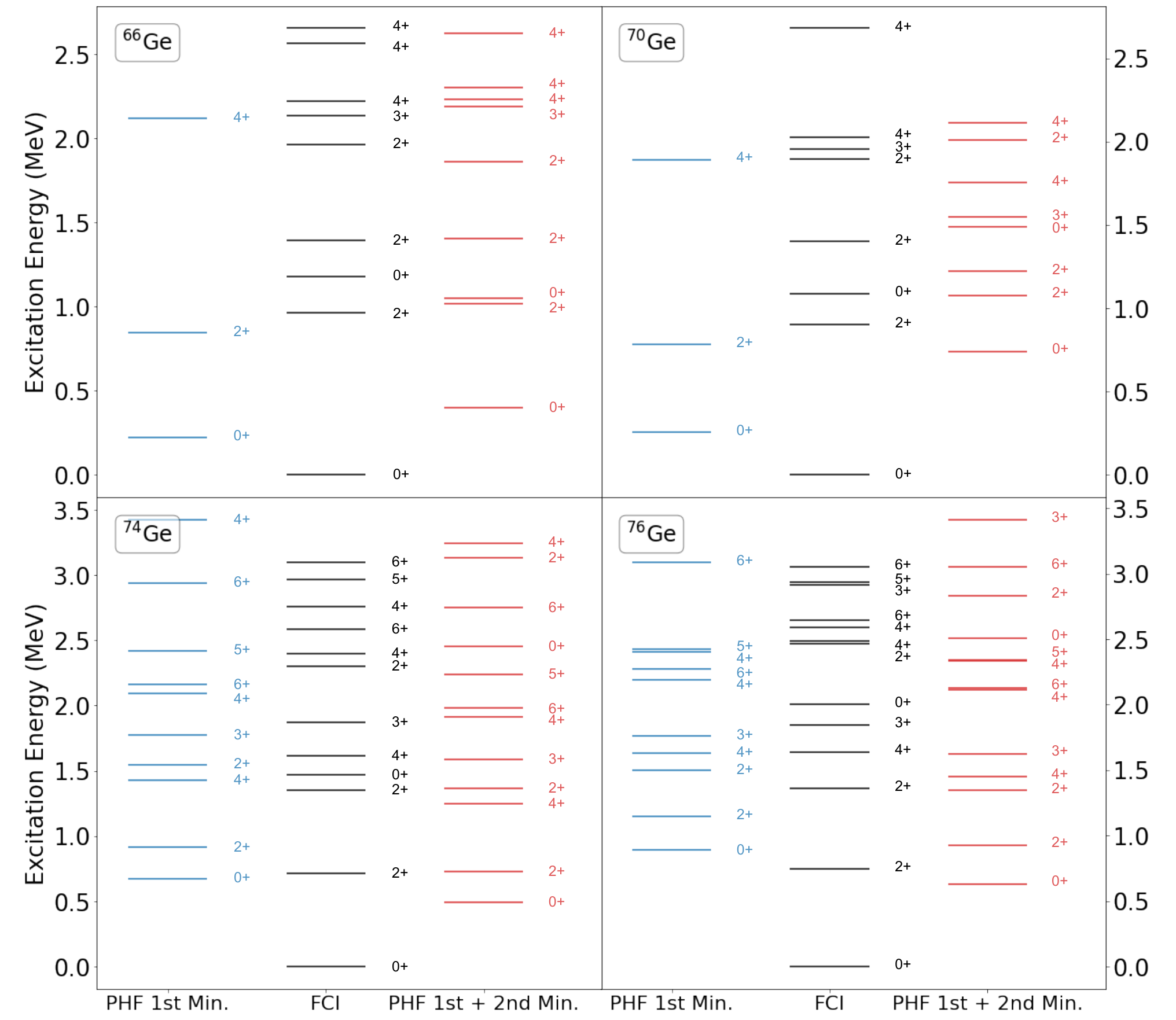}
    \caption{Same as Fig.~\ref{fig:sdShell}, but for selected even-$A$ germanium isotopes.
All calculations were performed using the JUN45 interaction \cite{PhysRevC.80.064323}, in the $f_{5/2}pg_{9/2}$ model space.}
    \label{fig:Ge_evenA}
\end{figure}


Germanium nuclei, known for their complex spectra arising from triaxiality and shape coexistence \cite{PhysRevC.25.2812,PhysRevC.76.034317}, offer a strong test of  PHF in capturing excitation spectra.

 Table~\ref{tab:ge} gives the quadrupole shape deformation parameters 
for 
two shell-model interactions for the $1p$-$0f_{5/2}$-$0g_{9/2}$ space, JUN45 \cite{PhysRevC.80.064323}, and a modified \cite{PhysRevC.76.054312} version 
of the JJ44 interaction \cite{PhysRevC.70.044314}. These agree qualitatively  
well with density functional calculations \cite{PhysRevC.76.034317}, in particular 
 the SLy6 parameterization \cite{chabanat1997skyrme} of Skyrme Hartree-Fock calculations plus BCS pairing of a 
 zero-range pairing force~\cite{PhysRevC.60.034304}. 
 Here $\beta_1, \gamma_1$ refer to the 
shape parameters of the global Hartree-Fock, or first, minimum, while $\beta_2, \gamma_2$ are
the parameters for a secondary, shape coexistence minimum if it exists. $\Delta E_\mathrm{soln}$ refers 
to the difference in Hartree-Fock energies between the global minimum and the shape-coexistence solution. 
(Note that for the SLy6 calculation of $^{76}$Ge, the order of the solutions is switched due to pairing, 
not included here.) 
In several cases different forces have similar solutions but in
different order, e.g., for $^{70}$Ge, the global HF minimum for JUN45 is oblate with a 
triaxial shape coexistence solution, while for JJ44 and SLy6 the global minimum is triaxial and SLy6 has an 
oblate shape coexistence solution. 
While the  agreement is not exact, the overall congruence nonetheless is noteworthy.    {For the even-even 
germanium isotopes, $\Delta E_\mathrm{RMS}=0.38$ MeV; note that the the spectra in 
Fig.~\ref{fig:Ge_evenA} is compressed compared to our $sd$- and $pf$-shell examples.}

As with the $sd$- and $pf$-shells,  Fig.~\ref{fig:Ge_evenA} presents several excitation spectra of even-$A$ germanium isotopes, using the JUN45 interaction. 
Although the model 
space allows for unnatural parity states, our HF solutions for even-$A$ only had natural parity, 
and so we do not discuss the unnatural parity states in the spectra. 
Like our $pf$-shell examples, the $2_1^+$ from projection is systematically too low in excitation energy, or, arguably, 
the $0_1^+$ ground state is not low enough.   {This latter interpretation is bolstered by allowing $E_\mathrm{shift}$ in Eq.~(\ref{dErms}) to float when minimizing $\Delta E_\mathrm{RMS}$, leading to 
the PHF ground state appearing 0.5 MeV or more above the FCI ground state.
}



Both $^{66,70}$Ge have oblate global HF minima with simple rotational excitation 
spectra; the mixing of the second, triaxial local minima in each case puts the $0^+_2$ and other 
features in roughly the right place, albeit lower than in FCI. The $^{66}$Ge PHF spectrum is better than that of 
$^{70}$Ge.  

The spectra from the triaxial global HF minima only for $^{74,76}$Ge are more complex, and mixing in 
the second local HF minima do little more than put a second $0^+$ at approximately the right excitation energy; 
overall the agreement between PHF and FCI is poorer.

Even with the inclusion of stable shape coexistence states, the PHF results for odd-$A$ germanium isotopes did not reflect the  complex excitation spectra, and so we do not present them. 



Overall, even with the inclusion of stable shape coexistence states, PHF excitation spectra do not perform as well 
in approximating the FCI spectra of germanium isotopes.  One can turn this around:  
germanium is a stringent test of any beyond-mean-field methodology. 

\subsection{Separation energies and odd-even staggering}

\label{separation}

\begin{figure}[h!]
    \centering
    \includegraphics[width=0.9\textwidth]{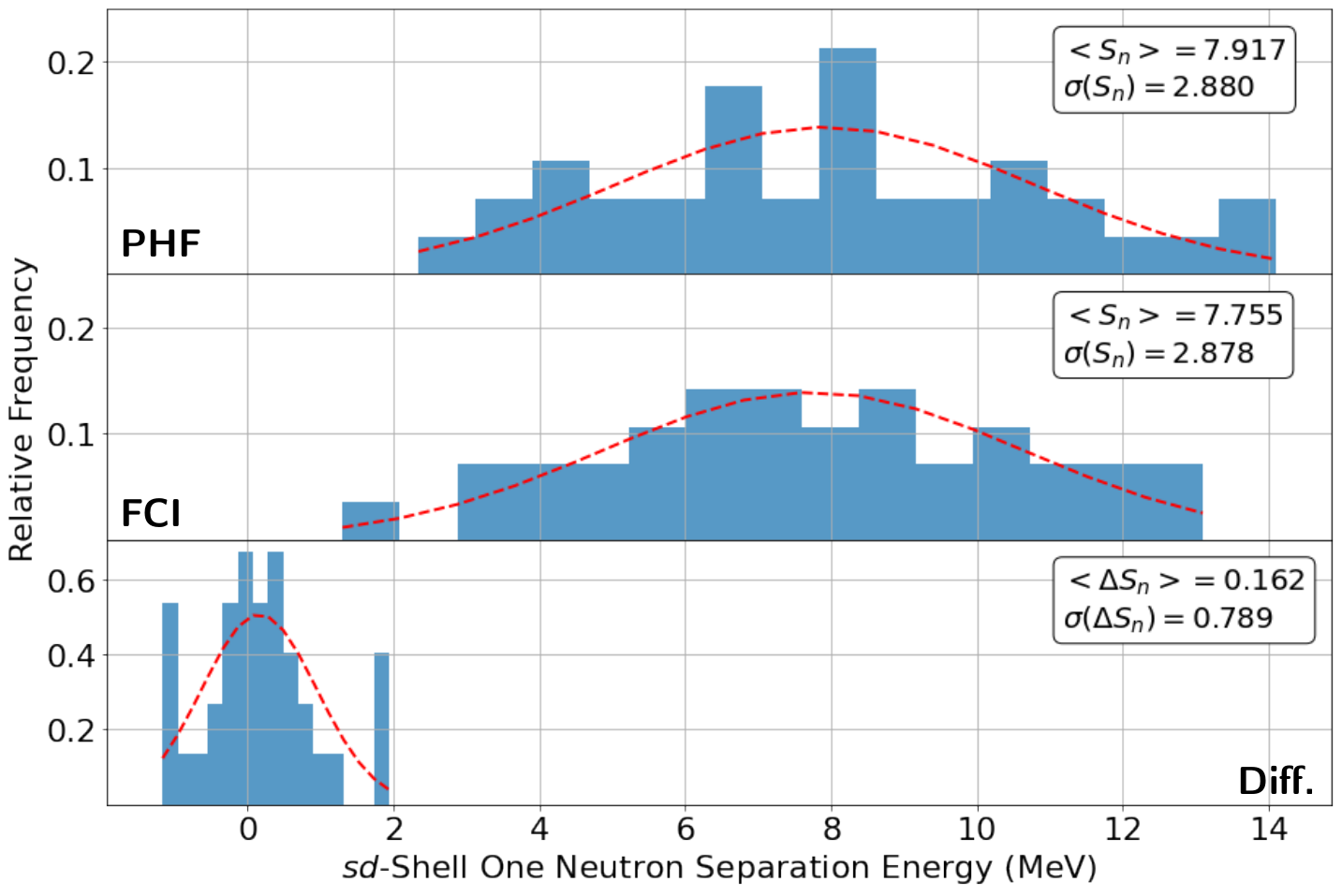}
    \caption{Distribution of neutron separation energies in the $sd$-shell computed 
    with FCI and  PHF, and the difference between the two methods. 
    }
    \label{fig:oneNsep}
\end{figure}

 As PHF provides a relatively simple ground state wave function, one cannot expect it to reproduce the same ground state binding 
 energy as FCI. Hence our focus so far on excitation energies. Nonetheless, one can also inquire about 
 relative binding energies. Here we consider the neutron separation energies, $S_n = BE(Z,N)-BE(Z,N-1)$,
 and the neutron odd-even staggering (OES)
~\cite{PhysRevLett.81.3599,PhysRevC.79.034306},
 \begin{equation}
\Delta_o^{(3)}(N) = \frac{1}{2}     \left [ BE(Z,N+1) + BE(Z,N-1) - 2 BE(Z,N) \right ]
 \end{equation}
 where $BE(Z,N)$ is the binding energy for a given $Z,N$.  Both the neutron separation energy 
 and the odd-even staggering is expected to be sensitive to  pairing, and could be 
  tests of  how the number-conserving PHF approximation may fail.
On the other hand, odd-even staggering occurs in random interactions even with the explicit pairing 
matrix elements set to zero~\cite{PhysRevC.61.014311}, so such an interpretation is not ironclad.

A comparison of the  neutron separation energy of the PHF and FCI results show good agreement between the data sets in the $sd$-shell. Figure \ref{fig:oneNsep} shows the  distributions of the PHF and FCI results as well as the  distribution of the differences.

\begin{figure}[h!]
    \centering
    \includegraphics[width=0.9\textwidth]{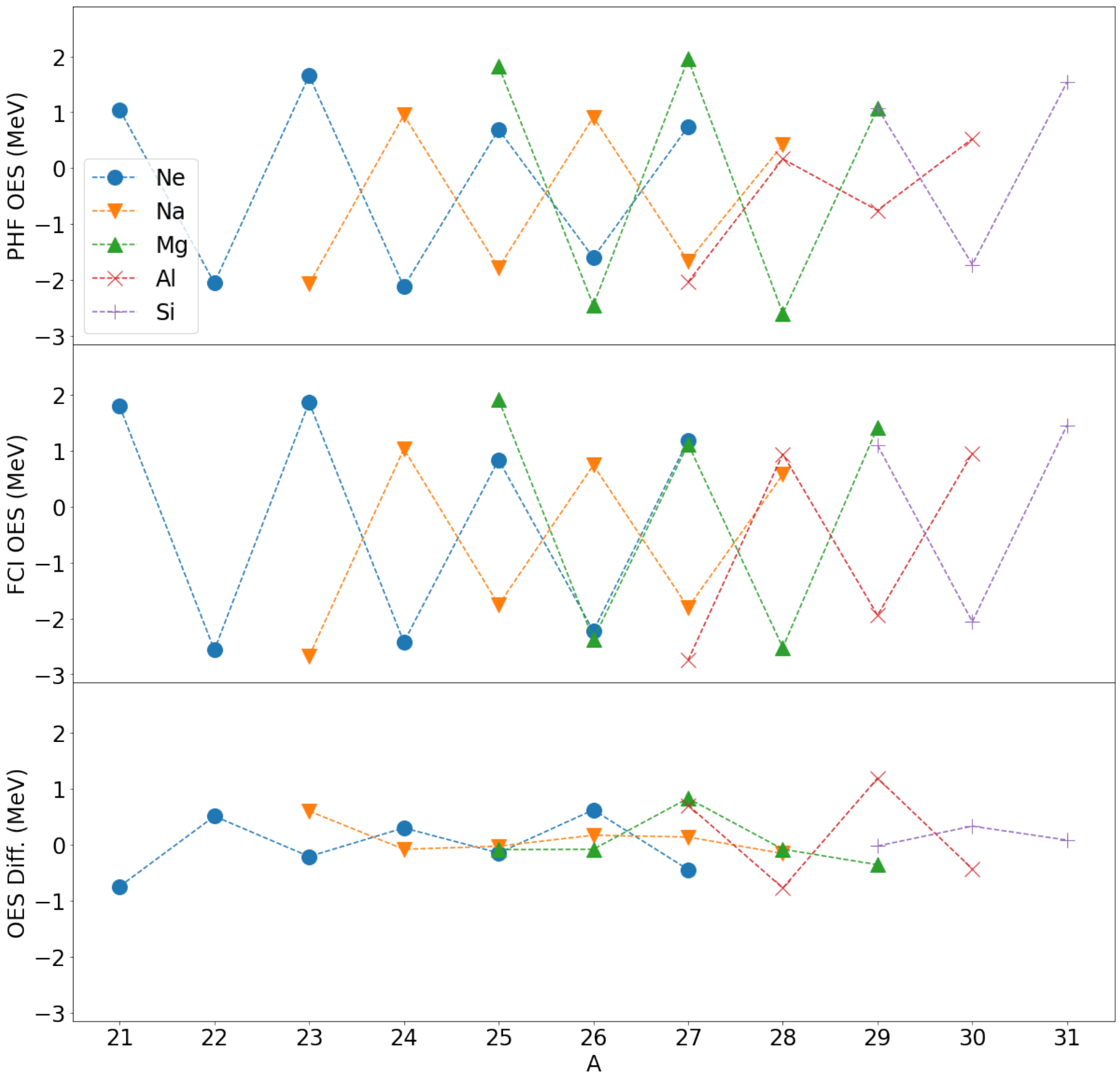}
    \caption{Odd-even staggering (OES) in $sd$-shell nuclei. 
    }
    \label{fig:oddEvenDiff}
\end{figure}

The $sd$-shell OES for PHF and FCI is shown in Figure \ref{fig:oddEvenDiff} with the difference between the two methods shown on the same scale in the bottom panel.
The average of the magnitude for the OES for PHF and FCI are 1.483 MeV and 1.672 MeV, respectively, with
the average magnitude of the difference only 0.378 MeV.  {We found that computing the OES with unprojected Hartree-Fock 
energies produced very poor results. We also found no strong evidence for shape staggering~\cite{PhysRevLett.81.3599}.}

\begin{figure}
    \centering
    \includegraphics[width=0.9\textwidth]{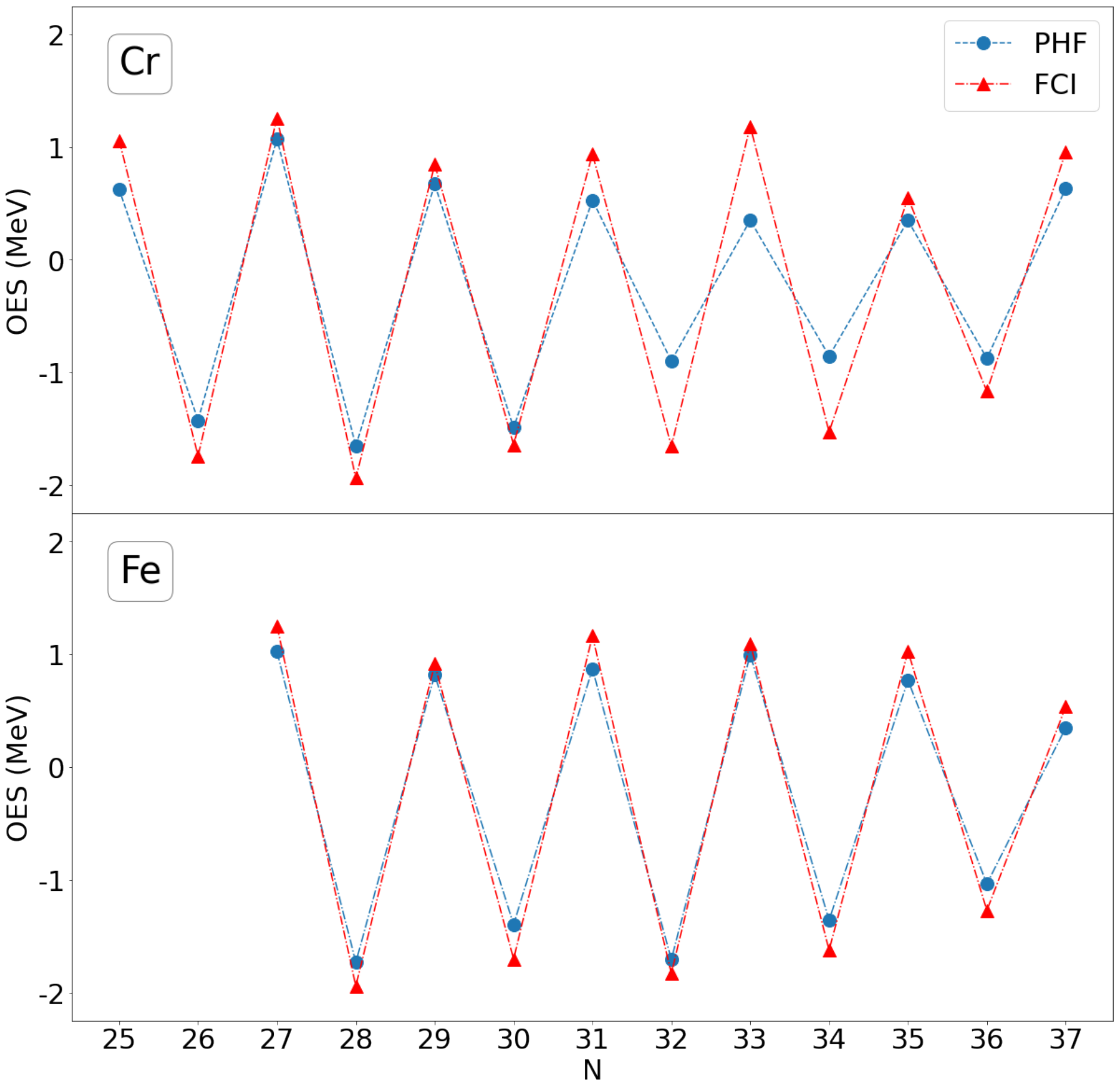}
    \caption{Odd-even staggering (OES) along the chromium (upper panel) and iron (lower panel) isotopic chains.}
    \label{fig:pfShellOES}
\end{figure}

\begin{figure}
    \centering
    \includegraphics[width=0.9\textwidth]{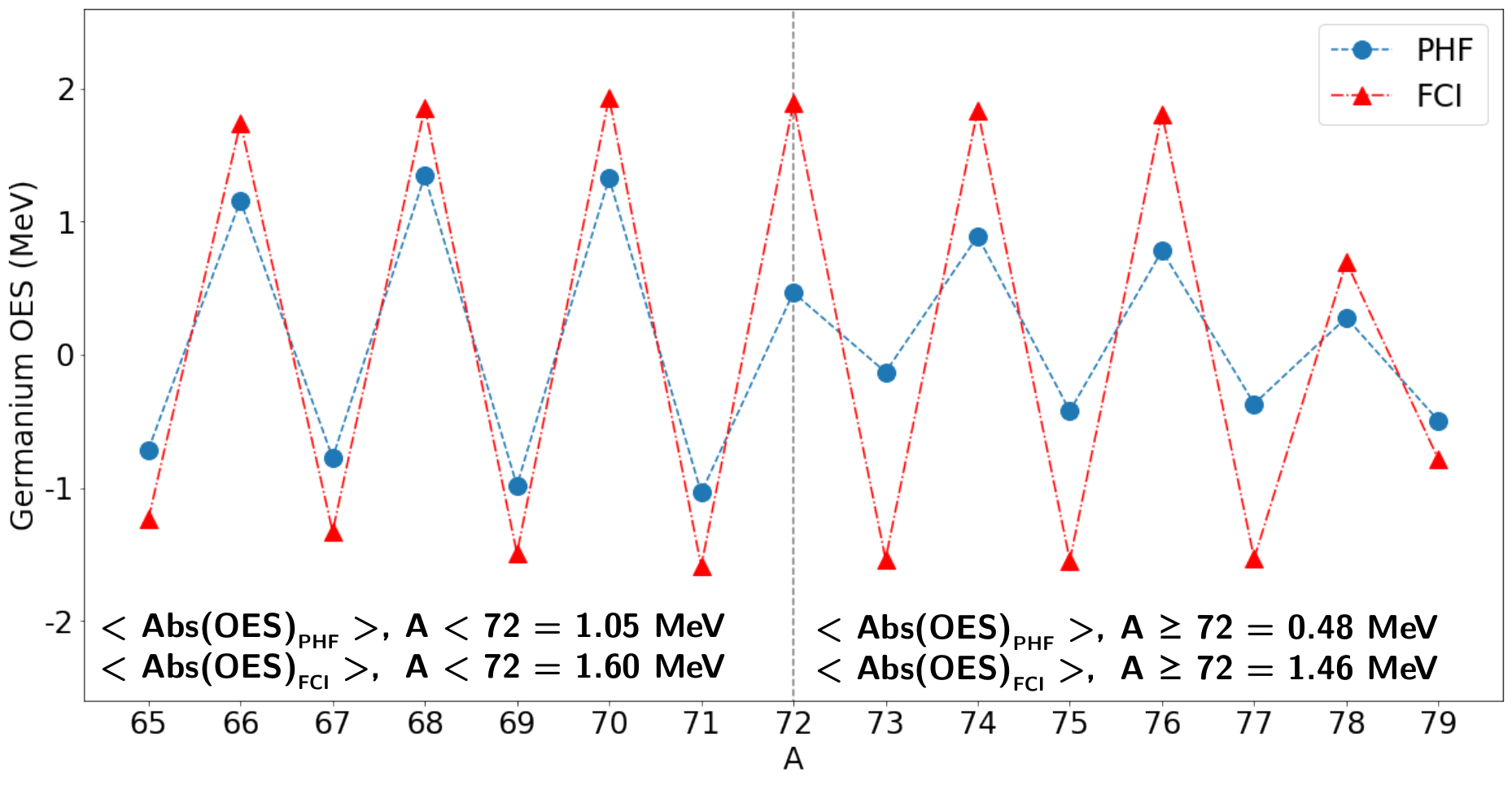}
    \caption{Odd-even staggering (OES) along the germanium isotopic chain.
    }
    \label{fig:Ge_OES}
\end{figure}

We also computed the OES for chromium and iron isotopes in the $pf$-shell, 
Fig.~\ref{fig:pfShellOES}, and for
germanium isotopes, in Fig.~\ref{fig:Ge_OES}. The former does very well.
Interestingly, for the latter, we obtain reasonable 
agreement for $A < 72$, with the PHF OES about two-thirds of the FCI values, but for $A \geq 72$, the PHF values 
are about a third of the FCI results. As this transition occurs at $N=40$ as one enters the $0g_{9/2}$ shell, 
one plausible explanation is the lack of good treatment of pairing. 

 {To test the role of pairing, we recomputed the OES for the chromium isotopic chain with increased pairing. 
Specifically, we added $-0.5 \hat{P}^\dagger \hat{P}$, where $\hat{P}^\dagger$ is the standard pair-creation operator,
$\hat{P}^\dagger = \sum_{j, m>0} \hat{a}^\dagger_{j,m} \hat{a}^\dagger_{j,-m} $. This roughly doubles the strength of the $J=0,T=1$ matrix elements found in GX1A. The deformations were unchanged, but the PHF OES was roughly half that of the FCI OES. 
This provides further evidence that weak pairing in these nuclides allows a good reproduction of differences in binding energies. In heavier nuclei where pairing becomes more important, PHF may not work as well; but there it is more challenging to 
carry out systematic FCI calculations for benchmarking. }


\subsection{Comparison to experimental $^{40}$Mg results}

\begin{figure}[h!]
    \centering
    \includegraphics[width=0.9\textwidth]{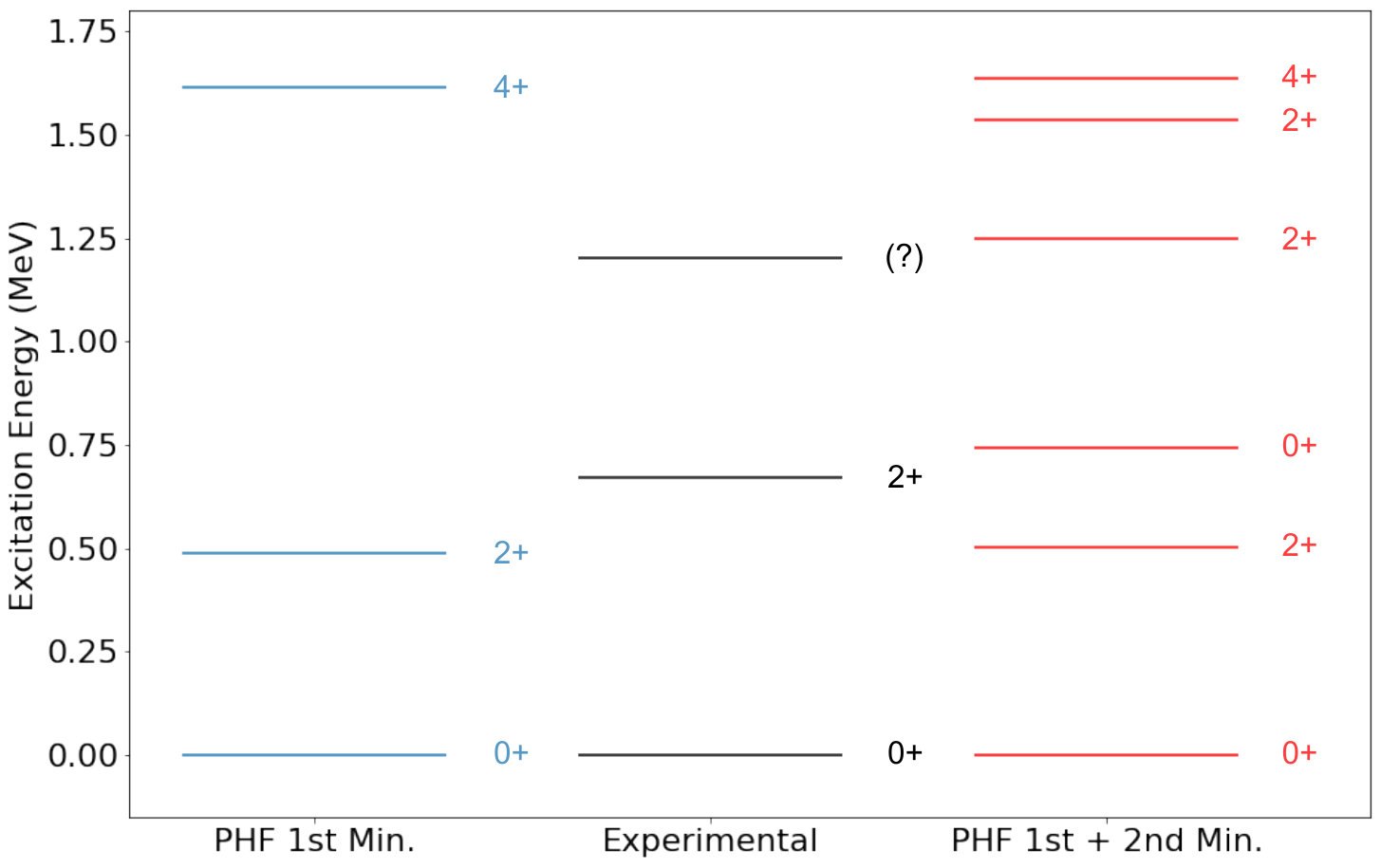}
    \caption{PHF calculation of $^{40}$Mg compared to experimental results from Crawford et al.\cite{crawford2019discovery}.}
    \label{fig:Mg40}
\end{figure}

Finally, to demonstrate the utility of simple PHF calculations in a shell model basis, 
we consider the neutron-rich nucleus $^{40}$Mg, recently measured \cite{crawford2014motivation,crawford2019discovery}.
A calculation in the $sd$-$pf$ shell using
a monopole-modified universal interaction \cite{PhysRevC.86.051301}, would require an $M$-scheme bais dimension 
of $10^{11}$, an order of magnitude larger than currently attainable. We compare the experimental results 
against PHF calculations in Fig.~\ref{fig:Mg40}. We produce the first excited $2^+$ state at roughly the 
right location, albeit a little low, consistent with our other calculations.  The $J^\pi$ of the second excited 
state has not been determined, but in our calculation it seems likely the $0_2^+$ is the head of a second band arising 
from a second minimum. 


\section{Conclusions and acknowledgements}

We have compared  {angular-momentum} projected{-after-variation} Hartree-Fock spectra to full configuration-interaction calculations, carrying out both in the same shell-model framework, using the same interaction matrix elements. 
PHF, also in use in quantum chemistry \cite{jimenez2012projected}, is a first step towards more sophisticated approximations, such as (projected) Hartree-Fock-Bogoliubov, 
generator coordinate, and the Monte Carlo shell model. Yet already, for a relatively simple and fast approximation 
PHF often provides qualitative agreement with FCI excitation spectra; key to this agreement are 
 general deformations, including triaxiality, and the use of gradient descent to find the HF minima.
Mixing in a second local minima when it exists, an additional step towards the more complex methods, 
often significantly improves the spectra.  {For our calculations the rms error in the spectra is around 0.78 MeV, 
with odd-$A$ performing significantly worse (rms error of about 1 MeV) than for even-even and odd-odd (rms error 
around 0.7 MeV for the $sd$- and $pf$-shells, and around $0.4$ MeV for germanium isotopes).} 

An important concern is pairing.   We have two contradictory pieces of information. The first is that for even-even 
nuclides we
tend to get the $2_1^+$ too low, or relatively speaking the ground state $0^+$ is not low enough, and this 
trend increases as we move up in shells. Here the lack of treating pairing is a viable suspect. 
 {While one might expect pairing to also affect neutron separation energies and the odd-even staggering of binding energies,
PHF provides a reasonable approximation to FCI calculations. When we increase the strength of pairing in chromium 
isotopes, however, the PHF estimate of the OES badly underperforms. }

While methods to compute the overlap of Bogoliubov vacua have gotten more efficient~\cite{PhysRevLett.108.042505,carlsson2020new}, 
it is nonetheless time-consuming to project out states of good particle number as well as good angular momentum. 
Given that the simpler PHF already encompasses many features of low-lying nuclear spectra, it is reasonable to 
ask whether simple configuration mixing of a few states, in the direction of generator coordinate methods or 
the Monte Carlo shell model, could compete with projected HFB. This question we leave to future work.  We also find complex germanium spectra a challenge to reproduce, 
even when using shape coexistence. Thus germanium provides a stringent test of 
any beyond-mean-field method.

This material is based upon work supported by the U.S. Department of Energy, Office of Science, Office of Nuclear Physics, 
under Award Number  DE-FG02-03ER41272, and by the Office of High Energy Physics, under Award No.~DE-SC0019465.

\bigskip

\bibliographystyle{unsrt}
\bibliography{johnsonmaster}

\end{document}